\documentclass{llncs}

\usepackage{color,latexsym,amssymb,epsfig,epic,eepic}

\newcommand{\Poly}{{\rm Poly}}

\begin{document}

\title{About the Lifespan of Peer to Peer Networks
\thanks{The research was partially funded by the European projects COST Action 293,
``Graphs and Algorithms in Communication Networks'' (GRAAL) and COST
Action 295, ``Dynamic Communication Networks'' (DYNAMO).}
\thanks{ Special thanks go to Jean-Claude Bermond. The research was
done while the authors were meeting at the MASCOTTE project site in
Sophia Antipolis.}}

\author{Rudi Cilibrasi\inst{1}, Zvi Lotker\inst{1}, Alfredo Navarra\inst{2}, Stephane
Perennes\inst{3} and Paul~Vitanyi\inst{1} \institute{ {CWI -
Kruislaan 413, NL-1098 SJ Amsterdam, Netherlands.\\ Email: {\tt
\{Rudi.Cilibrasi, Z.Lotker, Paul.Vitanyi\}@cwi.nl}} \and
{LaBRI - Universit\'e de Bordeaux 1, 351 cours de la Liberation,\\
33405 Talence, France. Email: {\tt Alfredo.Navarra@labri.fr}}\and
{MASCOTTE project, I3S-CNRS/INRIA/Universit\'e de Nice \\ Sophia
Antipolis, France. Email: {\tt Stephane.Perennes@sophia.inria.fr}}}}

\maketitle
\date{}

\begin{abstract}
In this paper we analyze the ability of peer to peer networks to
deliver a complete file among the peers. Early on we motivate a
broad generalization of network behavior organizing it into one of
two successive phases. According to this view the network has two
main states: first centralized - few sources (roots) hold the
complete file, and next distributed - peers hold some parts (chunks)
of the file such that the entire network has the whole file, but no
individual has it. In the distributed state we study two scenarios,
first, when the peers are ``patient'', i.e, do not leave the system
until they obtain the complete file; second, peers are ``impatient''
and almost always leave the network before obtaining the complete
file.

We first analyze the transition from a centralized system to a
distributed one. We describe the necessary and sufficient conditions
that allow this vital transition. The second scenario occurs when
the network is already in the distributed state. We provide an
estimate for the survival time of the network in this state, i.e.,
the time in which the network is able to provide all the chunks
composing the file. We first assume that peers are patient and we
show that if the number of chunks is much less than $e^n$, where $n$
is the number of peers in the network, then the expected survival
time of the network is exponential in the number of peers. Moreover
we show that if the number of chunks is greater than $\frac{\log
n}{n+1}e^{n+1}$, the network's survival time is constant. This
surprisingly suggests that peer to peer networks are able to sustain
only a limited amount of information. We also analyze the scenario
where peers are impatient and almost always leave the network before
obtaining the complete file. We calculate the steady state of the
network under this condition. Finally a simple model for evaluating
peer to peer networks is presented.
\end{abstract}


\section{Introduction}
Over recent years, peer to peer (P2P) networks have emerged as the
most popular method for sharing and streaming data (see for
instance~\cite{STYSIM05}). There has been popular adoption and
widespread success due to the high efficiencies that these networks
obtain for broadcast data. Apart from personal usage, many
companies, like for instance \textbf{red}hat$_{\tiny
\textregistered}$, provide links in order to download their free
distributions in a P2P fashion. In doing so, companies avoid the
problem of too many clients connected to their server. This solves
bottleneck or high concentrated transmission cost on a single node
with a significant chance of failure at peak loads. On the other
hand, companies are not the only benefactor. Indeed, it makes things
much faster from the point of view of the user even though at the
expense of ``being used'' by other users.

Another very interesting application where such networks are highly
successful concerns the distribution of data for storage purposes.
The idea, in fact, to collect data among users spread over all the
world is increasing more and more. Instead of having (for each one)
a full copy of everything, a community can share resources hence
obtaining a distributed storage device. This permits them to
collectively maintain more and more data and it increases also the
reliability. It ensures, in fact, that data will not disappear due
to the malfunction of a small number of devices.

The main differences among the two applications we have just
outlined are a more collaborative environment and a lower percentage
of disappearing pieces of data, with the second being the more reliable in
this sense. For downloading purposes, in fact, the aim is sometimes to download
the required data as fast as possible and then leave the network. This
implies also a higher frequency in peers disappearing. On the other
hand, in an ideal world where people collaborate for a common final
purpose, we like to imagine both the applications are quite
equivalent.

We consider the following processes. A file is divided into $k$
chunks. The network contains a large number of nodes. We distinguish
among {\em peers}, i.e., nodes with a number of chunks less than $k$
and {\em roots}, i.e., nodes owning all the chunks. We assign to
peers a probability $\alpha$ for which they may disappear. This
means that peers live on average $\frac{1}{\alpha}$ rounds. For the
roots we chose a probability $\alpha_R \geq \alpha$. We consider
closed network, i.e., every time a peer or a root leaves the network
a new peer will join with no chunks. For the sake of simplicity we
consider a synchronous model. During a round each node can receive
or send one chunk; and at the end of each round any node disappears
with respect to the related probability. We study the following
three scenarios:

\

\textbf{Spreading or Centralized Scenario:} Peers contain no chunks
and there are $R$ roots. We wonder if the {\em file is spread into
the network}. This happens if all the chunks are sent from the roots
inside the network where they multiply themselves. This can give us
a measure for the file length with respect to the life of the last
surviving root. After the last root leaves the network, in fact,
either the file has been spread or it is not possible to build it
back. We say in this case, the network (or the file) is ``dead''.

\

\textbf{Distributed Scenario:} The chunks are widely spread in the
network. There is not a fixed amount of roots, $R$ can be also zero,
we only require that the whole chunks composing the file are there.
We wonder if the network life is long or short. As we are going to
see, the network life is long when some almost steady state is
reached. This can give us a measure about the conditions ensuring a
long life for the network, and, when this happens, how often a full
download is completed.

\

\textbf{Survivability Scenario:} In this scenario we are interested
in studying the network behavior under extreme conditions. We
consider, in fact, the case in which a file is almost never
downloaded since peers have a very high volatility. They almost
never stay in the network long enough to perform a full download.
However, it is still very interesting to note that the network is
able to survive. We recall that a network is said to survives
whenever it still contains all the $k$ chunks composing the original
file, no matter where they reside.

\

Given some parameter settings, our aim is to answer the question of
how long we can expect a network to continue producing new completed
downloads. For all the previous three scenarios we provide a
stochastic formulation. We show how parameters should be set up in
order to obtain the desired results concerning network survivability
and file downloading rate. It turns out that the eventual fate of
the network is mainly dependent on the number of nodes $n$ and the
number of chunks $k$ in which the file is split. We show how the
network may pass through the previous three scenarios before
eventually dying.

\subsection{Related Work}
A lot of work has been devoted to the area of file sharing in P2P
networks. Many experimental papers provide practical strategies and
preliminary results concerning the behavior of these kind of
networks. In~\cite{IUBFAG04} for instance, the authors essentially
describe properties like liveness and downloading rate by means of
extended experiments and simulations under several assumptions.

Concerning analytical models it is very difficult to capture
suitable features in order to describe what happens and why
protocols like BitTorrent~\cite{C03} are so powerful in practice.
Suitable models are hard to find that describe what sometimes is
easily observable by simulations. One of the main assumption made in
the literature in order to describe the behavior of such networks at
a top level concern Flow models~\cite{CN04,QS04}, Queueing
theory~\cite{GFJKT03}, Network Coding~\cite{JLC05} and Coupon
Collector aspects~\cite{MV05}. This latter paper mainly focuses on
systems in which peers owning some chunks (usually one at random)
appear in the network with some probability and disappear as soon as
they complete their download. Recently in~\cite{AP06}, the
distribution of $k$ chunks on a network with diameter $d$ and
maximum degree $D$ has been proved to require at most $O(D(k + d))$
rounds of concurrent downloads with high probability. This is tight
within a factor of $D$. They also specialized to the networks used
by BitTorrent improving the bound to $O(k \ln \ n)$ rounds where $n$
is the number of nodes.

\subsection{A first thought}
Such results are quite interesting from a theoretical point of view
but sometimes not truly representative of the real life. The main
assumption that collides with practical aspects is that the number
of peers participating in the protocol is assumed to be huge, hence
obtaining asymptotically optimal results in terms of network
survivability and spreading speed of the desired file. Moreover,
different from our model many of the previously cited papers do not
assume the possibility for a peer to leave the network before it
completes the whole file. This aspect is indeed introduced also
in~\cite{QS04}. On the other hand we should immediately point out
that if there is at least one root that stays indefinitely, then the
file will always be available in the network, if a peer is willing
to wait. We call such a scenario ``{\em trivial}'' since there is no
question about the behavior of the network. In contrast, if all the
original $R$ roots disappear (at some time $t$) then there are many
possibilities. We say a chunk is present in the network if at least
one peer has that chunk. If there exists a chunk that is not present
on the network, then no more full downloads are possible. Therefore,
the most interesting case is when no roots are persistent ($t <
\infty $).

When using BitTorrent~\cite{C03} or similar programs in order to
download desired files, usually such networks look quite
different. The assumption for which a huge number of nodes is
participating in the protocol given in~\cite{JLC05,MV05,QS04} is
indeed too strong. Moreover, in practice, the number of chunks is
usually much bigger than the number of nodes composing the network.
This is due to the fact that even if a file is spread among
thousands of users, they do not participate concurrently in the
protocol. At any given time the network is usually quite small if
compared to the actual number of downloads.

\begin{figure}[h]
\begin{center}
\psfig{figure=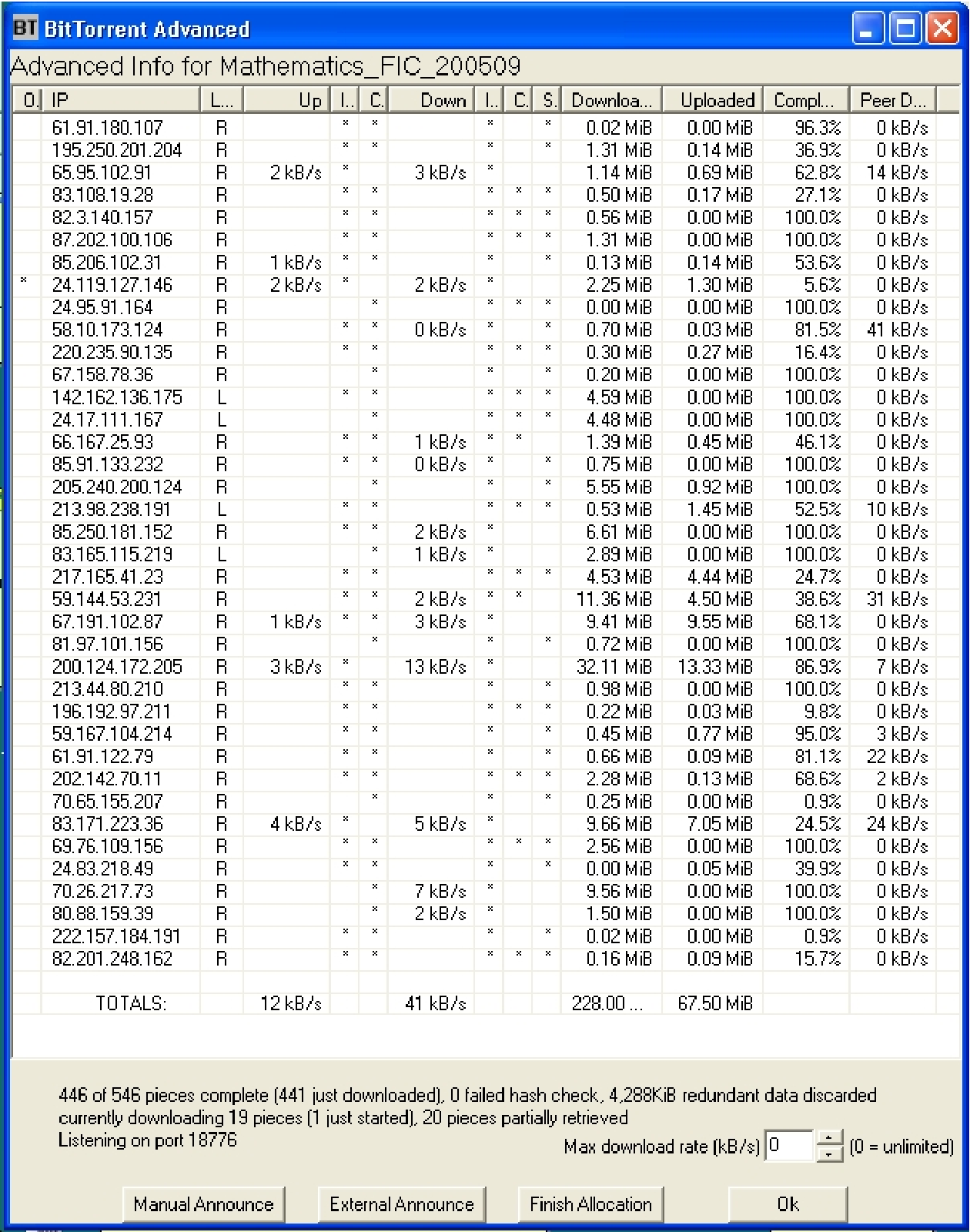,scale=0.32}\caption{Screenshot of the advanced
properties of BitTorrent during a downloading phase.\label{AdBit}}
\end{center}
\end{figure}

Figure~\ref{AdBit} shows a standard screenshot of the advanced
BitTorrent window while downloading a file of size roughly $272Mb$.
As it is described in the figure, there are $38$ peers participating
in the protocol with $15$ roots and $23$ peers while the number of
chunks is $546$ that is $512Kb$ per chunk\footnote{Indeed in the
BitTorrent specifications~\cite{C03}, the default size of a chunk is
$256Kb$, hence obtaining $1092$ chunks.}. It is worth noting that
under those circumstances the success of these protocols has to
reside in the adopted strategies, in contrast with that outlined
in~\cite{MV05}. In such a setting, for instance, the ``rarest
pieces'' distribution becomes quite important. This is the
peculiarity in BitTorrent for which once an empty peer appears in
the network it is provided with the least common chunk among the
network. Also the ``altruistic user behavior'' is quite crucial from
the point of view of the network survivability. It is based on the
observable fact for which peers that terminate their download do not
immediately disappear. In~\cite{IUBFAG04} for instance, it is
pointed out how friendly users usually behave in the network. They
do not disappear as soon as their download is finished thus ensuring
that all the chunks are available. Indeed most of the time this
happens since whoever is downloading has just left the computer
unattended but working.

\subsection{Outline}
The remainder of the paper is organized as follows. In the next
section we give our first insight of P2P networks by introducing the
so called Spreading Scenario. We show under which conditions a file
is successfully spread over the network, hence remains alive.
Section~\ref{sc22} describes the so called Distributed Scenario. We
show under which conditions a file spread over the network can
survive according to the number of peers composing the network and
the number of chunks in which the file is split. Section~\ref{sc2}
is devoted to the so called Survivability Scenario. In this case the
behavior of the network is studied under critical circumstances like
the high volatility of peers persistence. Section~\ref{model}
provides a simple model for P2P networks in order to obtain
numerical results about peers volatility, chunks distributions and
downloading rate. Finally, Section~\ref{concl} provides some
conclusive remarks.

\section{Spreading Scenario}\label{sc1}
In this section we study our first scenario in which a file must be
distributed among the network by spreading its $k$ chunks. Our model
is synchronous. At round $t$ the network is composed by $R_t$ roots
that disappear with probability $\alpha_R$. In this scenario we do
not take into account the number of peers. Usually peers are much
more than roots but for our analysis we just consider that a root
can provide one chunk to one single peer at each round. This can be
seen also by considering at round $t$ a number of peers equal to
$R_t$ since we are just analyzing the number of rounds needed by the
roots in order to spread the file. At a generic round $t$ each peer
asks a root for one random chunk. A root answers with a random chunk
that has never been spread inside the network before. This
implements the previously described ``rarest pieces'' issue. After
each round, roots coordinate with each other in order to maintain
the list of chunks that have been sent across the network.

The process is then similar to a coupon collector
problem~\cite{MV05}. All the $k$ chunks have to be collected, but a
chunk can be collected several times during a round. Moreover the
number of roots from which one can collect a coupon decreases
exponentially. Let $K_t$ be the number of chunks to be collected at
time $t$, i.e., chunks that have not been distributed until round
$t$.

For a given chunk $x$:
\[ Pr( x \mbox{ is  not collected} ) = \left(1-\frac{1}{K_t}\right)^{R_t}=
\left(1-\frac{1}{K_t}\right)^{ K_t \frac{R_t}{K_t}} \approx
e^{-\frac{R_t}{K_t}}\]

hence $E[K_{t+1} |K_t] \approx e^{-\frac{R_t}{K_t}} K_t$. Assuming
for now that $K_t=E[K_t]$ with probability $1$ we get $E[K_{t+1}]
\approx e^{-\frac{R_t}{K_t}} E[K_t]$.

For $R_t$ the situation is simpler since $R_t$ is just the sum of
$R$ independent variables, (each one being described by the series
$\sum Pr(R_{t+1}= i) z^i =  ( \alpha_R +  z (1-\alpha_R))^{R_t}$,
and $R_t$ is concentrated around its mean $(1-\alpha_R) R_{t}$. So
we have $E[R_{t+1}] =  \alpha_R E[R_t]$.

Let $\rho_t= E[R_t]/E[K_t]$, we get
\[ \rho_{t+1}= (1-\alpha_R) e^{\rho_t} \rho_t. \]

From this, two situations can follow. If $(1- \alpha_R ) e^{\rho_0}
\geq 1$, then $\rho_t$ always increases, and this increase is faster
and faster. This implies that the spreading will easily succeed
since the number of chunks not spread decreases much faster than the
number of roots that leave the network. Conversely, if $(1-
\alpha_R) e^{\rho_0} < 1$, $\rho$ decreases and keep doing it faster
and faster. This means that the process dies soon.

In the first situation we almost always collect all the chunks
otherwise never. Of course, either at some point the chunks are all
distributed or there are no roots that can provide the missing
chunks. We can conclude from this first analysis that the file gets
spread whenever $\alpha_R < 1 - e^{-\frac{R}{k}}$. Note that when
$k>>R$ this actually means $\alpha_R < \frac{R}{k}$ and this is
usually the case. For $k=R$ we get $\alpha_R \leq \frac{e-1}{e}$
but, in real world scenarios, this usually does not happen.

\section{Distributed Scenario}\label{sc22}
In the previous section we gave a necessary and sufficient condition
to describe the asymptotic behavior of the network. That is, the
network must move from the initial state to the distributed state.
In this section we study the behavior of P2P networks in the
distributed state, that is, there are no roots, yet every chunk is
available on the network after time $t$. In the following we refer
to~\cite{W91} for the applied probabilistic tools.

\subsection{Upper bound }\label{Upper bound}
Our next goal is to show that networks that are in the distributed
state will not survive if the number of chunks is exponentially big
in the number of peers. In order to show this we assume the
following model. In each time step each peer asks a random chunk
among all the chunks that the peer is missing. We assume that if the
chunk is anywhere on the network then the peer will get this chunk
in the next time step. Clearly this assumption is optimistic and
will help the survival of the network. Importantly, this makes the
network's gross behavior deterministic and thus we can say with
certainty that every peer stays precisely $k$ timesteps before
leaving, since he gets exactly $1$ chunk per timestep. When a new
peer enters, it has no chunks and we may use the variable $i$, with
$0 \le i < n$, to indicate its ordering when all peers are sorted
according to their number of chunks. This ordering is equivalent to
the chronological ordering. After $\frac{k}{n}$ time steps, each
peer is promoted to the next ordinal position. We point out that
nothing changes in the network viability unless a peer leaves, and
further the only peer that may leave is the last one, or the one
with the most chunks; eventually the last peer will have $k$ chunks
when the file has been completely downloaded. We wish to bound the
probability that a chunk will be missing. Therefore the number of
chunks that each peer $i$ has is $\frac{k(i-1)}{n}$ whenever a peer
is leaving.

To be precise, we imagine the total network state at any time to be
given by a binary vector of length $kn$; that is,
$$\Omega = \left( \{ 0,1 \}^{k} \right)^{n}.$$
In $\Omega$, the first $k$ coordinates describe the chunks held by
the first peer. In this first part, the first coordinate is $1$ if
and only if the first peer holds the first chunk, otherwise, it is
$0$. The next coordinate indicates the next chunk, and so on for all
$k$ chunks.

We will define an event $G$ on $\Omega$ such that $\forall i=1..n$,
peer $i$ has exactly $\frac{ik}{n}$ chunks. Let $A^j_i$ be the event
that peer $i$ has chunk $j$. Let $X^j_{i}$ be the indicator variable
of $A^j_i$. Let $Y^j=\max\{X^j_1,X^j_2...,X^j_n\}$, i.e., $Y^j$ is
the indicator variable of the event that chunk $j$ is in the system.
We define the random variable $Z=\min\{Y^1,...,Y^k\}$ as follows: it
is the indicator variable of the event that there is a missing chunk
in the network. In other words, $Z=0$ means the network has died,
and $Z>0$ means it continues to distribute the file.

\begin{lemma}Let $n$ and $k$ be the number of peers and chunks in the system respectively. For
$n>2$, the probability that there is a missing chunk can be bounded
by $\Pr[Z=0]\leq k n e^{-n}$.
\end{lemma}

\begin{proof}
by the Union bound over all $k$ chunks,
$$\Pr[Z=0]<k \Pr[Y^1=0] = k \prod_{i=1}^{n} \frac{i }{n}= k \frac{n!}{n^n}$$
and by Sterling's approximation,
$$k \frac{n!}{n^n} <k n e^{-n}.$$
$\hfill\square$
\end{proof}

The next corollary shows that if the number of chunks is small the
probability for the network to die approaches 0.
\begin{corollary} For all $k < \frac{e^n}{n \log n}$,
$$\lim_{n\rightarrow \infty} \Pr[Z=0] =0.$$
\end{corollary}

The next corollary shows that if the number of chunks is small the
system survives for a long time.

\begin{corollary}
If the number of chunks is a polynomial $\Poly(n)$ then the expected
survival time is at least $\frac{e^n}{n \Poly(n)}.$
\end{corollary}

\subsection{Lower bound }\label{lower bound}
The main problem in proving the lower bound is the dependence
between the random variables $Y^j$ and $Y^{j'}$. To remove this
technical difficulty we use a different model, the binomial model.
The idea is to make $Y^j$ and $Y^{j'}$ i.i.d. variables. In order to
prove a lower bound on the previous model we increase the expected
number of chunks that peer $i$ has at the time the last peer leaves
the network. I.e., we relax the assumption that, at the time the
last peer leaves the network, the number of chunks in the peer $i$
is $\frac{k (i-1)}{n}$. Moreover we assume that the number of chunks
is a binomial random variable. This assumption is legitimate since
the binomial distribution is highly concentrated. The problem with
this approach is that now we are no longer sure that each peer has
enough chunks. The way we solve this problem is by strengthening the
peer capabilities by increasing the chance that a peer has received
chunk $i$. Since in a normal file sharing system chunks are
correlated and peers have a smaller number of chunks, our lower
bound also captures the behavior of these systems. This is justified
since both assumptions (increasing the number of packets and the
fact that packets are i.i.d) decrease the probability of failure. We
do not offer a method to achieve this, but rather we use this
approach to prove a lower bound with high probability. We posit this
property (the binomial distribution) for the proof. We assume that
the peers $i=1,...,n$ have $\frac{k i}{n+1}$ chunks on average. More
precisely, let $\mathcal{X}^j_{i}$ be a Bernoulli random variable
such that $E[\mathcal{X}^j_{i}]=\frac{i}{n+1}$. Let
$\mathcal{G}_i=\sum_{j=1}^k \mathcal{X}^j_{i}$ be a random variable
that counts the number of chunks that peer $i$ has. Note that
$E[\mathcal{G}_i]=\frac{k i}{n+1}$. The next lemma bounds the
probability that peer $i$ will have less than $\frac{(i-1) k}{n}$
chunks.

\begin{lemma}
\label{rlem} Let $\mathcal{G}_i$ be the number of different chunks
belonging to peer $i$. For all $ 1 \le i \le n$,
$\Pr[\mathcal{G}_i<\frac{(i-1) k}{n}]< e^{-\frac{ k}{2 n^3(n+1)} }$.
\end{lemma}

\begin{proof}
$$\Pr\left[\mathcal{G}_i<\frac{k (i-1)}{n}\right]=\Pr\left[\mathcal{G}_i<
\left(1-\frac{n+1-1}{i n}\right)\frac{ki}{(n+1)}\right]<$$
$$<e^{-\big(\frac{n+1-i}{in}\big)^2 \frac{k i}{2 (n+1)}}<
e^{-\frac{k}{{2 n^3(n+1)} }}$$ The first equality follows from
algebra, and a Chernoff bound yields the next inequality.
$\hfill\square$
\end{proof}

If $k >> n^4$ we get that the probability that the $i$-peer
(Bernoulli process) has less than $\frac{(i-1)k}{n}$ chunks is
exponentially small.

Let $Q=\bigcap_{i=0}^{n-1} \{\mathcal{G}_i\geq\frac{(i-1)k}{n} \}$.
Note that $Q$ is the event that all peers have more chunks than they
are supposed to have, i.e., for all $i$,
$\mathcal{G}_i\geq\frac{(i-1)k}{n}$.

\begin{lemma}
For all $\log k>n $, $\Pr[Q]> 1-  n e^{-\frac{ k}{2(n+1)n^3}}$.
\end{lemma}

\begin{proof}
$$\Pr[Q] = \prod_{i=1}^{n} \Pr\left[\mathcal{G}_i\geq \frac{k
(i-1)}{n}\right] =$$ $$=\prod_{i=1}^{n}
\left(1-\Pr\left[\mathcal{G}_i < \frac{k (i-1)}{n}\right] \right)
\geq \prod_{i=1}^{n} \left (1-e^{-\frac{ k}{2(n+1)n^3} }\right ).
$$

We apply Lemma~\ref{rlem} to derive the last inequality above. We
choose the smallest term in the product and raise it to the $n$
power for the bound:

$$\Pr[Q]\geq \left (1-e^{-\frac{  k}{2(n+1)n^3} }\right )^{n}
> 1-n e^{-\frac{ k}{2(n+1)n^3} }.$$ $\hfill\square$
\end{proof}

Using the previous lemma it follows that the probability for which
$\overline{Q}$ holds is exponentially small.

\begin{corollary} For all $\log k>n$,
$\Pr[\overline{Q}]< n e^{-\frac{ k}{2(n+1)n^3} }$.
\end{corollary}

After bounding the probability that all the Bernoulli peers will
have more chunks than the discrete peers, we analyze the probability
that the Bernoulli peers will fail, i.e., some chunk is missing. Let
$\mathcal{Y}^j=\max\{\mathcal{X}^j_1,\mathcal{X}^j_2...,\mathcal{X}^j_n
\}$, $\mathcal{Z}=\min\{\mathcal{Y}^1,...,\mathcal{Y}^k\}$. Note
that $\mathcal{Z}=0$ is equivalent to say that the Bernoulli peers
will fail.

The following lemmata lead to the last corollary that shows under
which condition the network goes to miss some chunk, i.e., it is not
able to deliver any further complete download.

\begin{lemma}
\label{ylem}The probability that the Bernoulli peers will fail is,
$$\Pr[\mathcal{Y}^j=0]>\frac{1}{e^{n+1}}.$$
\end{lemma}

\begin{proof}
The proof follows from the following computation.
$$\Pr[\mathcal{Y}^j=0]=\prod_{i=1}^n \frac{i+1}{n+1}=\frac{(n+1)!}{(n+1)^n}>
\frac{\frac{(n+1)^{n+1}}{e^{n+1}}}{(n+1)^{n+1}}= \frac{1}{e^{n+1}}.$$
$\hfill\square$
\end{proof}

\begin{lemma}
$$\Pr[\mathcal{Z}>0]<e^{-\frac{k(n+1)}{e^{n+1}}}.$$
\end{lemma}

\begin{proof}
In order to prove the claim we make use of Lemma~\ref{ylem},
followed by the limit definition of $e$ and then Sterling's
approximation, hence obtaining
$$\Pr[\mathcal{Z}>0]=\prod_{j=1}^k \left(1-\Pr[\mathcal{Y}^j=0]\right)=
\left(1-\frac{(n+1)!}{(n+1)^n}\right)^k \cong
e^{-\frac{k(n+1)!}{(n+1)^n}}<e^{-\frac{k(n+1)}{e^{n+1}}}.$$
$\hfill\square$
\end{proof}


\begin{lemma}
For $\log k >n$,
$$\Pr[Z=0]\geq 1-e^{-\frac{k(n+1)}{e^{n+1}}}- n e^{-\frac{ k}{2(n+1)n^3}}.$$
\end{lemma}

\begin{proof}
From Bayes Law and the complementary events property,

$$\Pr[Z=0]> \Pr[\mathcal{Z}=0| Q] = \Pr[\mathcal{Z}=0]-\Pr[\overline{Q}]
\Pr[\mathcal{Z}=0|\overline{Q}] >$$
$$>\Pr[\mathcal{Z}=0]-\Pr[\overline{Q}] \geq
1-e^{-\frac{k(n+1)}{e^{n+1}}}- n e^{-\frac{ k}{2(n+1)n^3}}.$$
$\hfill\square$
\end{proof}

\begin{corollary} For all $k\geq \frac {\log n}{n+1}e^{n+1}$,
$$\lim_{n\rightarrow \infty} \Pr[Z=0] =1.$$
\end{corollary}

From the previous corollary it follows that if the number of chunks
$k$ is bigger than or equal to $\frac {\log n}{n+1}e^{n+1}$, then
the expected survival time is constant.

\section{Survivability Scenario}\label{sc2} In this section we study how
likely the network is to survive in extreme conditions. With extreme
conditions we mean that a file is almost never downloaded since
peers have a very high volatility and almost never stay in the
network long enough to perform a full download. However, it is still
very interesting to note that the network is able to survive. We
remind that a network survives whenever it still contains all the
$k$ chunks composing the original file, no matter where they reside.
We will assume that peers leave the system with probability $\alpha$
while roots or {\em experts} (peers that succeed at the full
download) leave the network with probability $\alpha_R$.

During the process chunks are duplicated and hence created while
others disappear because of nodes leaving the system. Let us denote
by $N$ the number of nodes (peers plus experts, assuming roots as
experts) in the network and by $P_i$ the probability (percentage of
peers) that a peer has $i$ chunks. The number of chunks lost during
one time step is then:

\[N \left(\alpha \sum_{i=0}^{k-1}  i P_i  + \alpha_R P_k \right).\]

The amount $C$ of chunks created, depends on how many successful
download are performed during a step. This strongly depends on the
chosen protocol. In any case this amount cannot be more than the
number of peers with one packet $N(1-P_0)$ multiplied by the
probability to stay inside the network, i.e., $C \leq
(1-\alpha)N(1-P_0)$.

Hence a necessary condition for the network survival is that

\[  N(1-P_0) (1-\alpha)  \geq  N\left(\alpha \sum_{i=0}{k-1} i P_i
+ \alpha_e P_k \right) \geq  N \alpha (1-P_0)\]

and this leads to have $1-\alpha \geq \alpha$, i.e., $\alpha \leq
\frac{1}{2}$.

When $\alpha=\frac{1}{2}$ the stationary distribution is as follows
$P_0=P_1=\frac{1}{2}$, after each communication step all the nodes
have a chunk, but then half of them die (reset to zero chunks). Note
that such a network dies quite quickly, from deviations. However, as
soon as $\alpha > \frac{1}{2}$ the network lives almost forever.

Our process is very similar to a birth and death process, each node
lives on average $\frac{1}{1-\alpha}$ rounds. And at each round it
generates $1-\alpha$ chunks, hence its average number of children is
$\frac{1 -\alpha}{\alpha}$. It holds that when this number is
strictly greater than $1$ the network survives.

Let us assume the network to be in some random state with
$P_0=P_1=\frac{1}{2}$, with $k$ chunks regularly spread across the peers
with one chunk each. Let $F_{i,t}$ be a random variable that denotes
the percentage of peers with the $i$-th chunk at time $t$. $F_{i,0}$
is deterministic with value $\frac{N}{2k}$.

After the first step of the protocol, we have:

\[ P[ F_{i,1}= k] =  {2F_{0,t} \choose k } / 2^{2F_{i,0}}\]

So after the initial step, chunks are distributed as the sum of
$\frac{N}{2k}$ random bits.

We study this phenomenon at the critical point in its most canonical
form. At time $t$ we have a number  $S_t$ of chunks alive, we double
each chunk and randomly destroy half of the chunks.

First we consider the future of a single chunk. Let $F_t(z)$ denote
the generating series at time $t$ associated with the future of
chunk $z$.

Considering one time step we have: with probability $\frac{1}{4}$
the process dies, with probability $\frac{1}{2}$ the process
restarts with $1$ chunk and $t-1$ time units remain, with
probability $\frac{1}{4}$ we get two chunks and $t-1$ time units
remain.
\[ F_t(z) = \frac{1}{4} + \frac{1}{2} F_{t-1}(z) + \frac{1}{4}
F^2_{t-1}(z) = \left( \frac{ F_{t-1}(z) +1 }{2} \right)^2  \]

Note that by setting $\varepsilon_t(z)_t= F_t(z)-1$, we obtain

\[ \varepsilon_t(z)= \varepsilon_{t-1}(z) +\left(\frac{\varepsilon_{t-1}(z)}{2}\right)^2\]
and $\varepsilon_t(0)$ is the probability to stop before time $t$.
So the probability to have the process alive at time $t$ is about
$\frac{t}{\ln t}$, and if one considers $n$ bits one needs $n \ln n$
time units to kill all of them.

This scenario is quite optimistic since one assumes that the network
exchanges $N(1-P_0)$ chunks during a round which corresponds to a
{\em perfect} situation. This happens, in fact, by means of a {\em
perfect matching} between the nodes that meet their necessities.
Indeed, if we make use of a non optimal strategy, i.e., matching the
node in a perfectly arbitrarily way, it is worth noting that this
does not affect and degrade the process too much.

A typical way to marry the nodes is to choose randomly for each node
a server, and to elect a node randomly for each server. Despite the
simplicity of this process (nodes which have all the chunks still
demand some, and nodes with no chunks are considered as able to
provide some), it does not affect the network survivability as we
are going to see.

Note that, by means of a simple matching algorithm, the number of
edges is already of order $(1-\frac{1}{e}) N$. Among those edges
some are unable to duplicate chunks (if the server does not have any
chunks that the client needs). A critical stage occurs when almost
all the nodes with chunks have almost no chunks, so the probability
for an edge to be useful is indeed exactly $(1-P_0)$. In such a
situation the number of chunks replicated is
 \[ \left(1-\frac{1}{e} \right) \left(1 -P_0\right) \left(1-\alpha\right) \]

while the number of chunks destroyed is at least $\alpha(1-P_0)$.
This implies that the network survives when $\alpha \leq
\frac{1-\frac{1}{e}}{ 2 - \frac{1}{e}}\simeq .3873$.

\section{Simple evaluation Model for P2P Networks}\label{model}
To better understand the behavior of this kind of network we propose
a simplified model and protocol. Such a model can be easily applied
and modified in order to find preliminary results on P2P networks.
We have also compared our easy model's results to the outputs of
more sophisticated simulators. Even though the comparison is outside
the scope of this paper, it is worth mentioning that the deviation
from the simulations is negligible.

\begin{itemize}
\item[-]
At the beginning of each time slot, each peer chooses another peer
randomly. An inquired peer randomly selects one of its customers. We
call this customer {\em lucky}.  Next the uploading peer delivers to
its lucky customer a useful random chunk, i.e., a chunk that he has
and that this customer has not. The unlucky customers do not get any
chunk at this round.\\

\item[-]
In order to get a stable situation, each peer that disappears (with
probability $\alpha$) is immediately replaced by a new empty peer
(i.e., with no chunks).
\end{itemize}

The probability $luck$ that a customer gets lucky is indeed equal to
the proportion of customers served, which is the number of peers
having at least one customer. Hence, $luck=1 -\frac{1}{e}$.

We first compute the probability that a node with $i$ chunks gets a
new useful one. To do this we make a strong assumption that chunks
remains almost identically distributed, i.e., a random node with $i$
chunks contains a given chunk with probability $\frac{i}{k}$.

To get a chunk, a node needs first to be lucky, then the probability
that it gets a chunk depends on the number of chunks of its uploading peer.
Assume that a customer with $i$ chunks contacts an uploading peer with $j$
chunks, its probability to receive a chunk is
\[ \Delta_{i,j}= 1 - \frac{ { i \choose j }  } { { k \choose j}  } \]

Note that we use the standard convention ${i \choose j}=0 $ whenever
$j \geq i$, but in the case $i=j=0$ we have $\Delta_{i,j}=0$, so we
consider ${ 0 \choose 0 } =1$. It follows that a lucky node in state
$i$ that does not vanish (i.e., conditioned on all those events)
moves to states $i+1$ with probability

\[ \sum_{j= 0*}^k \Delta_{ij} P_j \]

From that it follows that a peer in state $i\neq 0$ moves to:

\begin{itemize}
\item[-] State $i+1$ with probability $ T_{i,i+1}=  (1-\alpha)luck \sum_{j= 0}^k
\Delta_{i,j} P_j $

\item[-] State $i$ with probability $T_{i,i}=  (1-\alpha)  - T_{i,i+1}$

\item[-] State $0$ with probability $\alpha$
\end{itemize}

To summarize we have: $$P_0 = \alpha\sum_{i=0}^k P_j +
\left(1-\alpha\right)\left(1- luck \sum_{j= 0}^k \Delta_{0,j}
P_j\right),$$

$\forall i>0,$ $$P_i = (1-\alpha) \left( P_i \left( 1 - luck
\sum_{j= 0}^k \Delta_{i,j}   P_j\right) + P_{i-1} \left( luck
\sum_{j= 0}^k \Delta_{i-1,j} P_j\right) \right).$$

For the aim of preliminary and experimental results such a model is
already enough in order to get an idea of the general behavior of
this class of networks.

\section{Conclusion}\label{concl}
In this paper we have studied the behavior of P2P networks. We have
considered three main scenarios. In the first one there are some
peers owning all the chunks (roots) composing a file and the aim is
to study the time required to ensure that every chunk is spread out
on the network. This is very important to understand since, of
course, it reflects the required behavior for peers that want to
share their information. We have shown that the success of the
spreading phase depends on two main parameters. Namely, the number
of roots in the network and the number of chunks in which the file
is divided. The probability $\alpha_R$ for which $R$ roots can leave
the network should be smaller than $1-e^{-\frac R k}$ where $k$ is
the number of chunks. In the second scenario, we have started with a
configuration in which many peers have subsets of the whole chunk
set and the aim is to study the probability for the network to
survive, i.e., every chunk must belong to some peer. This is also
very important since it gives a measure of the behavior that peers
should exhibit in order to maintain the viability of their download
and archival capabilities. We have shown that if $k$ is much less
than $e^n$, with $n$ being the number of peers in the network, the
expected survival time of the network is exponential in $n$.
Moreover, if the number of chunks is greater than $\frac {\log n}
{n+1} e^n+1$, the network survival time is constant. The third
proposed scenario concerns the critical setting for the peers in
terms of volatility. We have shown how under this setting the
network is still able to survive. Namely, our estimated maximum
value of the probability $\alpha$ for which a peer can leave the
network while guaranteing its survivability is $\alpha\leq .3873$.
From the point of view of experimental results, we have also
proposed a simple way for analyzing and modeling P2P networks.

Our study has raised many open questions that might be investigated
for further research. Many variations of our proposed models are
possible and interesting. An important issue, for instance, concerns
file sharing protocols that cope with security aspects. Deep
analysis of tit-for-tat strategies for avoiding the so called
free-riders problem is of primary interest to better understand the
success of these protocols (see~\cite{JA05} for preliminary
results). Free-riders are users that download files from the network
but do not share their own chunks. In BitTorrent, those kind of
users are allowed even though the performance of their downloads is
much slower than for ``friendly'' users.

\end{document}